\documentclass[12pt]{article}
\usepackage{amsmath,amssymb,amsfonts}
\usepackage{graphicx}
\usepackage[square, comma, sort&compress]{natbib}
\setcitestyle{numbers}
\def\br(#1,#2){\left\langle#1#2\right\rangle}
\def\sq(#1,#2){\left[#1#2\right]}
\def\s(#1,#2){s_{#1 #2}}
\def\t(#1,#2,#3){s_{#1 #2 #3}}

\begin{document}

\begin{titlepage}
\hspace*{\fill}\parbox[t]{3.5cm}
{
\today} \vskip2cm
\begin{center}
{\Large \bf Effective Field Theory\\\bigskip
for Nonstandard Top Quark Couplings} \\
\medskip
\bigskip\bigskip\bigskip\bigskip

{\bf Nicolas Greiner, Scott Willenbrock, and Cen Zhang}\\
\bigskip
Department of Physics, University of Illinois at Urbana-Champaign \\ 1110 West Green Street, Urbana, IL  61801
\end{center}
\begin{abstract}
We present an effective-field-theory calculation of the effect of a dimension-six operator involving the top quark on precision electroweak data via a top-quark loop.  We demonstrate the renormalizability, in the modern sense, of the effective field theory.  We use the oblique parameter $\hat U$ to bound the coefficient of the operator, and compare with the bound derived from top-quark decay.
\end{abstract}

\end{titlepage}

%\section{Introduction}\label{sec:intro}%

There are two ways to search for physics beyond the Standard Model (SM).  One way is to search directly for new particles.  The other is to search for the indirect effects that new particles might have on the known SM particles.  This could manifest itself as nonstandard interactions of the known particles, sometimes called anomalous couplings.

The modern approach to nonstandard couplings is effective field theory \cite{Weinberg:1978kz}.  In the effective-field-theory approach, the SM is regarded as the leading approximation at ``low'' energies, that is, at energies much less than the scale of new physics, $\Lambda$.  The new physics enters as a correction to this leading approximation, suppressed by inverse powers of $\Lambda$.  For most observables, the leading correction is suppressed by two inverse powers of $\Lambda$.  This corresponds to operators in the Lagrangian of dimension-six, in contrast to the SM Lagrangian, where all operators are of dimension four or less.

The effective-field-theory approach has a number of virtues:
\begin{itemize}
\item It is well motivated and provides guidance as to the most likely place to observe the indirect effect of new physics.
\item The known $SU(3)_C\times SU(2)_L\times U(1)_Y$ gauge symmetry of the SM is respected.
\item It is model independent, and general enough to accommodate all possible physics beyond the SM.
\item Radiative corrections due to SM interactions are calculable and unambiguous.
\item Radiative corrections due to dimension six operators are calculable and unambiguous.
\end{itemize}
The effective-field-theory approach incorporates everything we already know about nature at high energy, and allows us to entertain the possibility of new physics without disrupting what has already been established.

A number of papers have advocated the use of effective field theory for top quark physics \cite{Cho:1994yu,Grzadkowski:1995te,Grzadkowski:1997cj,Grzadkowski:2003tf,Grzadkowski:2004iw,Grzadkowski:2005ye,
Cao:2006pu,Cao:2007ea,Lillie:2007hd,Kumar:2009vs,AguilarSaavedra:2008gt,AguilarSaavedra:2008zc,
AguilarSaavedra:2009mx,AguilarSaavedra:2010rx,AguilarSaavedra:2010zi,Zhang:2010dr,
Degrande:2010kt}.
%\cite{Cho:1994yu}-\cite{Degrande:2010kt}
Those papers have considered the effect of dimension-six operators on the production and decay of the top quark.  However, the top quark also plays an important role as a virtual particle in precision electroweak physics.  Indeed, the correct range for the top-quark mass was anticipated by precision electroweak studies.  Now that the top-quark mass is accurately known from direct measurements, we can ask what the precision electroweak measurements have to say about the presence of dimension-six operators in loop diagrams involving the top quark.  Because of the last virtue listed above, this is a well-defined question with an unambiguous answer.

In this paper, we will focus on just one dimension-six operator that affects the top quark,
\begin{equation}
O_{tW} = (\bar q\sigma^{\mu\nu}\tau^It)\tilde\phi W_{\mu\nu}^I,
\end{equation}
where $W_{\mu\nu}^I$ is the $SU(2)_L$ field-strength tensor, $\phi$ is the Higgs doublet, $t$ is the right-chiral top quark, and $q$ is the left-chiral doublet containing top and bottom.\footnote{$\sigma^{\mu\nu}=\frac{i}{2}[\gamma^\mu,\gamma^\nu]$ is a tensor constructed from Dirac matrices, and $\tau^I$ are the $SU(2)_L$ Pauli matrices. The top-quark fields are mass eigenstates, and $\tilde\phi=\epsilon\phi^*$.} We chose this operator because it is the only one that contributes to the leading correction to the decay of the top quark to $W$ bosons of a given helicity.  Thus this operator can already be bounded from present data.  We calculate the contribution of this operator to precision electroweak data via a top-quark loop and compare the resulting bound on the coefficient of this operator with the bound obtained from top-quark decay.

\begin{figure}[htb]
\centering\includegraphics[width=10cm]{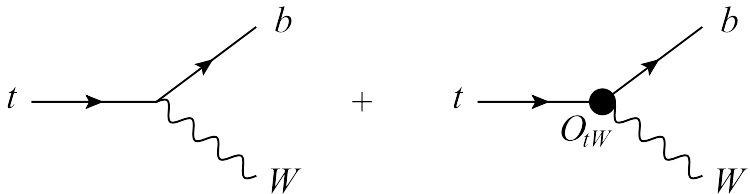}
\caption{The dimension-six operator $O_{tW}$ contributes to the top-quark decay process
through a correction to the $Wtb$ vertex.}\label{fig:fig1}
\end{figure}

When the Higgs field acquires a vacuum-expectation value, the dimension-six operator $O_{tW}$ yields the effective interactions \cite{AguilarSaavedra:2008zc}
\begin{eqnarray}
\mathcal{L}_{eff}&=&\mathcal{L}_{SM}
+\frac{C_{tW}}{\Lambda^2}\Big[(v(\bar{b}\sigma^{\mu\nu}(1+\gamma_5)t)\partial_\mu W_\nu^-+h.c.)
\nonumber\\&&+\sqrt{2}c_Wv(\bar{t}\sigma^{\mu\nu}t)\partial_\mu Z_\nu
+\sqrt{2}s_Wv(\bar{t}\sigma^{\mu\nu}t)\partial_\mu A_\nu
-\sqrt{2}igv(\bar{t}\sigma^{\mu\nu}t)W_\mu^+W_\nu^-+
\cdots\Big]
\label{eq:Leff}\end{eqnarray}
where $C_{tW}$ is a dimensionless coefficient, $v\approx 246$ GeV is the Higgs vacuum expectation value, and $s_W,c_W$ are the sine and cosine of the weak mixing angle.
The first term in the effective interactions modifies the top-quark branching ratios to zero-helicity, negative-helicity, and positive-helicity $W$ bosons (see Fig.~\ref{fig:fig1}) \cite{Zhang:2010dr},
\begin{eqnarray}
f_0&=&\frac{m_t^2}{m_t^2+2m_W^2}-\frac{4\sqrt 2C_{tW}v^2}{\Lambda^2}\frac{m_tm_W(m_t^2-m_W^2)}{(m_t^2+2m_W^2)^2} \label{eq:f0}\\
f_-&=&\frac{2m_W^2}{m_t^2+2m_W^2}+\frac{4\sqrt 2C_{tW}v^2}{\Lambda^2}\frac{m_tm_W(m_t^2-m_W^2)}{(m_t^2+2m_W^2)^2}\\
f_+&=&0
\end{eqnarray}
where we have neglected the bottom-quark mass throughout, which is an excellent approximation.

We compare with recent data from the CDF \cite{CDF} and D0 \cite{Abazov:2010jn} collaborations, which report a measurement of $f_0$ (with the constraint $f_+ =0$ imposed):
\begin{eqnarray}
&&f_0=0.62\pm 0.11 \,({\rm stat})\pm 0.06 \,({\rm syst})\quad({\rm CDF}) \;,\\
%&&f_0=0.669\pm 0.078 \,({\rm stat})\pm 0.065 \,({\rm syst})\quad({\rm D0}) \;.
&&f_0=0.735\pm 0.051 \,({\rm stat})\pm 0.051 \,({\rm syst})\quad({\rm D0}) \;.
\end{eqnarray}
These measurements are consistent with the SM prediction, at NNLO in QCD \cite{Czarnecki:2010gb},
\begin{equation}
f_0=0.687(5)
\end{equation}
where the uncertainty is primarily from the uncertainty in the top-quark mass.
Because we are using an effective-field-theory approach, we can consistently include both QCD radiative corrections and the correction due to the dimension-six operator, which is the second-to-last virtue listed above.
Comparing with data yields the constraints
\begin{eqnarray}
&&\frac{C_{tW}}{\Lambda^2}=1.10\pm2.06\ {\rm TeV}^{-2}\quad({\rm CDF})\;, \label{eq:CDF}\\
&&\frac{C_{tW}}{\Lambda^2}=-0.79\pm1.19\ {\rm TeV}^{-2}\quad({\rm D0})\;. \label{eq:D0}
\end{eqnarray}
The NLO QCD correction to the second term in Eq.~(\ref{eq:f0}) is also known \cite{Drobnak:2010ej}.
It increases the value of $C_{tW}/\Lambda^2$ by about 1\%, much less than the uncertainty in this quantity.

\begin{figure}[htb]
\centering\includegraphics[width=14cm]{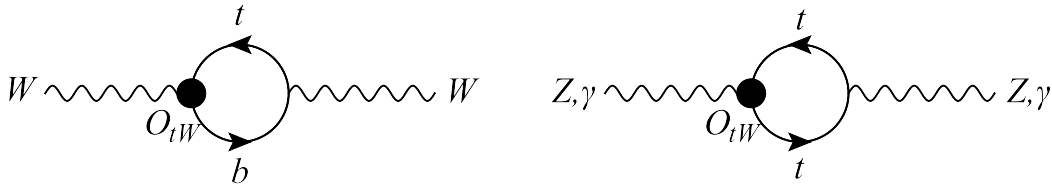}
\caption{The dimension-six operator $O_{tW}$ contributes to the electroweak-gauge-boson self energies via loop diagrams.}\label{fig:fig2}
\end{figure}

We now turn to the effect of $O_{tW}$ on precision electroweak measurements via a top-quark loop, as shown in Fig.~\ref{fig:fig2}.\footnote{There is also a diagram contributing to the $W$-boson self energy, with a top-quark loop, constructed from the contact interaction given by the last term in Eq.~(\ref{eq:Leff}).  Since this interaction is antisymmetric in $\mu,\nu$, this diagram does not contribute to the self energy.}  Since this loop only affects the electroweak-gauge-boson self energies, we may be able to use the well-known $S,T,U$ formalism to characterize it \cite{Peskin:1990zt,Peskin:1991sw,Barbieri:2004qk}.  Following Ref.~\cite{Barbieri:2004qk}, we define these oblique parameters in terms of self energies and derivatives of self energies at $q^2=0$,
\begin{eqnarray}
&&\hat S=-\frac{c_W}{s_W}\Pi^{\prime}_{30}(0)=c_W^2\Pi'_{ZZ}(0)-\frac{c_W}{s_W}(c_W^2-s_W^2)\Pi'_{\gamma Z}(0)-c_W^2\Pi'_{\gamma\gamma}(0)\label{eq:S}\\
&&\hat T=-\frac{\Pi_{33}(0)-\Pi_{11}(0)}{m_W^2}=\frac{1}{m_W^2}\left[\Pi_{WW}(0)-c_W^2\Pi_{ZZ}(0)\right]\label{eq:T}\\
&&\hat U=\Pi^{\prime}_{33}(0)-\Pi^{\prime}_{11}(0)=-\Pi'_{WW}(0)+c_W^2\Pi'_{ZZ}(0)+2c_Ws_W\Pi'_{\gamma Z}(0)+s_W^2\Pi'_{\gamma\gamma}(0)\;.\label{eq:U}
\end{eqnarray}
The contribution of the operator $O_{tW}$ to the oblique parameters, via Fig.~\ref{fig:fig2}, is calculated in dimensional regularization to be
\begin{eqnarray}
&&\hat S=N_c\frac{gC_{tW}}{4\pi^2}\frac{\sqrt{2}vm_t}{4\Lambda^2}\frac{5}{3}
\left(\frac{1}{\epsilon}-\gamma+\ln 4\pi-\ln\frac{m_t^2}{\mu^2}\right) \label{eq:Sloop}\\
&&\hat T=0 \label{eq:Tloop}\\
&&\hat U= N_c\frac{gC_{tW}}{4\pi^2}\frac{\sqrt{2}vm_t}{4\Lambda^2}\label{eq:Uloop}
\end{eqnarray}
where $N_c=3$ is the number of colors and $\mu$ is the 't Hooft mass.

\begin{figure}[htb]
\centering\includegraphics[width=6cm]{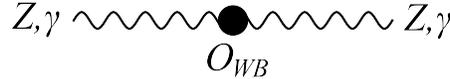}
\caption{The operator $O_{WB}$ contributes to the electroweak-gauge-boson self energies at tree level.}\label{fig:fig3}
\end{figure}

The contribution of $O_{tW}$ to the $\hat S$ parameter is ultraviolet divergent.  However, there is another dimension-six operator,
\begin{equation}
O_{WB}=(\phi^\dagger\tau^I\phi)W^I_{\mu\nu}B^{\mu\nu}\;,
\end{equation}
($B^{\mu\nu}$ is the $U(1)_Y$ field-strength tensor) that contributes to the $\hat S$ parameter at tree level, as shown in Fig.~\ref{fig:fig3}.  This operator must be included for consistency, since it also contributes to the $\hat S$ parameter at order $1/\Lambda^2$.  We find
\begin{equation}
\hat S=\frac{C^0_{WB}v^2}{\Lambda^2}\frac{c_W}{s_W}
\end{equation}
where $C^0_{WB}$ is the bare coefficient of the operator.
This coefficient is renormalized by the one-loop contribution of the operator $O_{tW}$ in Eq.~(\ref{eq:Sloop}).  In the $\overline{\rm MS}$ scheme, the total contribution to the $\hat S$ parameter is
\begin{equation}
\hat S = \frac{C_{WB}(\mu)v^2}{\Lambda^2}\frac{c_W}{s_W} - N_c\frac{gC_{tW}}{4\pi^2}\frac{\sqrt{2}vm_t}{4\Lambda^2}\frac{5}{3}
\ln\frac{m_t^2}{\mu^2}
\label{eq:Srenorm}\end{equation}
which is finite and unambiguous.  This is an example of the renormalizability of an effective field theory in the modern sense.  Although an effective field theory is not renormalizable in the old-fashioned sense, it is renormalizable at any order in $1/\Lambda^2$, provided that all the pertinent operators are included \cite{Gomis:1995jp}.

Although the result for the $\hat S$ parameter is finite and unambiguous, it cannot be used to constrain the coefficient $C_{tW}$, because of the tree-level contribution from the operator $O_{WB}$.  A measurement of the $\hat S$ parameter constrains only the linear combination of $C_{WB}$ and $C_{tW}$ contained in Eq.~(\ref{eq:Srenorm}).  For the choice $\mu=m_t$, a measurement of the $\hat S$ parameter constrains only $C_{WB}(m_t)$.

There is no contribution to the $\hat T$ parameter from the operator $O_{tW}$ [see Eq.~(\ref{eq:Tloop})].  Even if there were a contribution, there is also a tree-level contribution from the operator $O_{\phi}^{(3)}=(\phi^\dagger D^\mu\phi)[(D_\mu\phi)^\dagger\phi]$ that would mask the one-loop contribution from $O_{tW}$.  A top-quark model that gives a nonvanishing contribution to the $\hat T$ parameter is discussed in Ref.~\cite{Pomarol:2008bh}.

There is no tree-level contribution to the $\hat U$ parameter, defined by Eq.~(\ref{eq:U}), at order $1/\Lambda^2$, so the one-loop contribution from the operator $O_{tW}$, Eq.~(\ref{eq:Uloop}), is the sole contribution at this order.  The one-loop result is finite, as guaranteed by the renormalizability of the effective field theory in the modern sense.

The value of the $\hat U$ parameter may be obtained from Ref.~\cite{Nakamura:2010zzi}.  In Ref.~\cite{Nakamura:2010zzi}, the $U$ parameter is defined as
\begin{eqnarray}
\alpha U&=&
4s_W^2\left[\frac{\Pi_{11}(m_W^2)-\Pi_{11}(0)}{m_W^2}-\frac{\Pi_{33}(m_Z^2)-\Pi_{33}(0)}{m_Z^2}\right]
\nonumber\\
&=&4s_W^2\left[\frac{\Pi_{WW}(m_W^2)-\Pi_{WW}(0)}{m_W^2}
-\frac{c_W^2(\Pi_{ZZ}(m_Z^2)-\Pi_{ZZ}(0))+2s_Wc_W\Pi_{\gamma Z}(m_Z^2)+s_W^2\Pi_{\gamma\gamma}(m_Z^2)}{m_Z^2}\right]
\end{eqnarray}
($\alpha$ is the fine structure constant)
which apparently differs from the definition of $\hat U$ in Eq.~(\ref{eq:U}).  However, Ref.~\cite{Nakamura:2010zzi} tacitly assumes that the gauge boson self energies are linear in $q^2$, in which case the two definitions of $U$ are equivalent up to normalization: $\hat U = - \alpha U/4s_W^2$.  Nevertheless, we must also check whether our calculation of the contribution to the self-energy function from $O_{tW}$ is approximately linear in $q^2$.  Since the constraint on the $U$ parameter comes dominantly
from the measurement of the $W$-boson mass \cite{Nakamura:2010zzi},
it suffices to show that the linear approximation
is valid in predicting the value of $W$-boson mass.

In the $\hat{S},\hat{T},\hat{U}$ formalism, the $W$-boson mass can be expressed as
\cite{Peskin:1991sw}
\begin{eqnarray}
m_W^2&=&m_W^2(\mathrm{SM})
\left(
1-\frac{2s_W^2}{c_W^2-s_W^2}\hat{S}
+\frac{c_W^2}{c_W^2-s_W^2}\hat{T}-\hat{U}
\right)
\nonumber\\&=&m_W^2(\mathrm{SM})+
\frac{c_W^2}{c_W^2-s_W^2}\Pi_{WW}(0)+m_W^2\Pi'_{WW}(0)
\nonumber\\
&&-\frac{c_W^4}{c_W^2-s_W^2}\left[\Pi_{ZZ}(0)+m_Z^2\Pi'_{ZZ}(0)\right]+\frac{s_W^2c_W^2}{c_W^2-s_W^2}m_Z^2\Pi'_{\gamma\gamma}(0),\label{eq:mwlinear}
\nonumber\\\end{eqnarray}
where the definitions of $\hat{S},\hat{T},\hat{U}$ in Eqs.~(\ref{eq:S}-\ref{eq:U})
are used, and $m_W(\mathrm{SM})$ is the value
of the $W$-boson mass calculated as accurately as possible in the SM.

The exact formula for $m_W$, without assuming a linear dependence of the self energies on $q^2$, is
%\cite{Peskin:1991sw}
\begin{equation}
m_W^2=m_W^2(\mathrm{SM})+
\Pi_{WW}(m_W^2)+\frac{s_W^2}{c_W^2-s_W^2}\Pi_{WW}(0)-\frac{c_W^4}{c_W^2-s_W^2}\Pi_{ZZ}(m_Z^2)
+\frac{s_W^2c_W^2}{c_W^2-s_W^2}m_Z^2\Pi'_{\gamma\gamma}(0)\;.
\label{eq:mwexact}\end{equation}
Comparing Eqs.~(\ref{eq:mwlinear}) and (\ref{eq:mwexact}), we find that the error introduced by the linear approximation is
\begin{equation}
\delta m_W^2=
-\left[\Pi_{WW}(m_W^2)-\Pi_{WW}(0)-m_W^2\Pi'_{WW}(0)\right]
+\frac{c_W^4}{c_W^2-s_W^2}\left[\Pi_{ZZ}(m_Z^2)-\Pi_{ZZ}(0)-m_Z^2\Pi'_{ZZ}(0)\right]
\;.
\label{eq:deltamw}\end{equation}
For the operator $O_{tW}$, we find
\begin{eqnarray}
\delta m_W^2&=&
-N_c\frac{gC_{tW}}{4\pi^2}\frac{\sqrt{2}vm_t}{\Lambda^2}
m_W^2
\left\{
\frac{3-8s_W^2}{3(1-2s_W^2)}c_W^2\left(1-\frac{\sqrt{4m_t^2-m_Z^2}}{m_Z}\arctan\frac{m_Z}{\sqrt{4m_t^2-m_Z^2}}\right)
\right.\nonumber\\&&\left.
+\frac{1}{2}\left[\frac{m_t^2}{m_W^2}+\left(\frac{m_t^2}{m_W^2}-1\right)^2\ln\left(1-\frac{m_W^2}{m_t^2}\right)\right]-\frac{3}{4}
\right\}
\nonumber\\
&=&0.47\ \mathrm{GeV}^{2}\frac{C_{tW}}{\Lambda^2}\mathrm{TeV}^2\;.
\label{eq:error}\end{eqnarray}

Using the world-average $W$-boson mass, $m_W=80.399\pm 0.023$ GeV, the uncertainty in $m_W^2$ is $\delta m_W^2\approx 4$ GeV$^2$.  As we will see shortly, the value of $C_{tW}/\Lambda^2$ extracted from precision electroweak data is of order 1 TeV$^{-2}$, so the error introduced by the linear approximation, Eq.~(\ref{eq:error}), is an order of magnitude less than the experimental uncertainty in $m_W^2$. Thus the linear approximation is excellent, and we may use the $U$ parameter to bound
$C_{tW}/\Lambda^2$.  The linear approximation is valid because the expansion parameter for the contribution of the operator $O_{tW}$ to the self energies (Fig.~\ref{fig:fig2}) is $q^2/m_t^2$, and this parameter is sufficiently small for the values $q^2=m_W^2,m_Z^2$ needed in Eq.~(\ref{eq:deltamw}).

The value of the $U$ parameter is \cite{Nakamura:2010zzi}
\begin{equation}
U=0.06\pm 0.10
\end{equation}
for $m_t=173.0$ GeV and $m_h=117$ GeV, although there is very little dependence on the Higgs mass.  This corresponds to
\begin{equation}
\hat U=(-5.0 \pm 8.4)\times 10^{-4}
\end{equation}
Using Eq.~(\ref{eq:Uloop}), we find the constraint
\begin{equation}
\frac{C_{tW}}{\Lambda^2}=-0.7\pm1.1\ {\rm TeV}^{-2}
\label{eq:CtW}\end{equation}
which is slightly stronger than the constraint from the measurement of top-quark decay, Eqs.~(\ref{eq:CDF}) and (\ref{eq:D0}).

Thus far we have assumed that $O_{tW}$, $O_{WB}$, and $O_{\phi}^{(3)}$ are the only nonvanishing dimension-six operators.  We can relax this assumption by including, along with $O_{tW}$, all dimension-six operators that contribute to the gauge-boson self energies at tree level, which includes $O_{WB}$ and $O_{\phi}^{(3)}$.  These are \cite{Grinstein:1991cd}
\begin{eqnarray}
&O_{WB}=(\phi^\dagger\tau^I\phi)W^I_{\mu\nu}B^{\mu\nu}\;,
\qquad&
O_{\phi}^{(3)}=(\phi^\dagger D^\mu\phi)[(D_\mu\phi)^\dagger\phi]\;,
\label{eq:oblique O1}
\\
&O_{DB}=\frac{1}{2}(\partial_\rho B_{\mu\nu})(\partial^\rho B^{\mu\nu})\;,
\qquad&
O_{DW}=\frac{1}{2}(D_\rho W^I_{\mu\nu})(D^\rho W^{I\mu\nu})\;.
\label{eq:oblique O2}
\end{eqnarray}
Such operators originate
whenever heavy fields directly couple only to the SM gauge fields and the Higgs doublet.
Such operators
are sometimes referred to as ``universal.''

Once these operators are included, the self energies are no longer approximately linear functions of $q^2$, since $O_{DB}$ and $O_{DW}$ generate terms proportional to $q^4$.
Therefore we need four additional oblique parameters, which correspond to the second order
derivatives of the four self energies with respect to $q^2$.
Along with $\hat S,\hat T,\hat U$, we will use the four additional oblique parameters defined in Ref.~\cite{Barbieri:2004qk}:
\begin{align}
V=&-\frac{m_W^2}{2}(\Pi''_{33}(0)-\Pi''_{11}(0))
=\frac{m_W^2}{2}\left[\Pi''_{WW}(0)-c_W^2\Pi''_{ZZ}(0)-2c_Ws_W\Pi''_{\gamma Z}(0)-s_W^2\Pi''_{\gamma\gamma}(0)\right]
\\
W=&-\frac{m_W^2}{2}\Pi''_{33}(0)=-\frac{m_W^2}{2}\left[c_W^2\Pi''_{ZZ}(0)+2c_Ws_W\Pi''_{\gamma Z}(0)+s_W^2\Pi''_{\gamma\gamma}(0)\right]
\\
X=&-\frac{m_W^2}{2}\Pi''_{30}(0)=\frac{m_W^2}{2}\left[c_Ws_W\Pi''_{ZZ}(0)-(c_W^2-s_W^2)\Pi''_{\gamma Z}(0)-c_Ws_W\Pi''_{\gamma\gamma}(0)\right]
\\
Y=&-\frac{m_W^2}{2}\Pi''_{00}(0)=-\frac{m_W^2}{2}\left[s_W^2\Pi''_{ZZ}(0)-2c_Ws_W\Pi''_{\gamma Z}(0)+c_W^2\Pi''_{\gamma\gamma}(0)\right]
\end{align}
%Note that the $\hat{S}$, $\hat{T}$ and $\hat{U}$ are related to the usual $S$, $T$ and $U$ parameters as: $\hat{S}=\alpha S/4s_W^2$, $\hat{T}=\alpha T$, $\hat{U}=-\alpha U/4s_W^2$.

At tree level, four of the seven oblique parameters receive a contribution from a dimension-six operator:
\begin{align}
\hat{S}=&C_{WB}\frac{c_W}{s_W}\frac{v^2}{\Lambda^2},\\
\hat{T}=&-C_{\phi}^{(3)}\frac{v^2}{2\Lambda^2},\\
W=&-2C_{DW}\frac{m_W^2}{\Lambda^2},\label{eq:W}\\
Y=&-2C_{DB}\frac{m_W^2}{\Lambda^2}.
\end{align}
The other three oblique parameters, $\hat{U}$, $V$, and $X$, are zero at tree level. Thus
the contribution to these parameters from $O_{tW}$ at one loop (Fig,~\ref{fig:fig2}) must be finite, as guaranteed by the renormalizability of the effective field theory in the modern sense. We find
\begin{align}
\hat{U}=&
%A=-N_c\frac{gC_{tW}}{4\pi^2}\frac{\sqrt{2}vm_t}{\Lambda^2}
N_c\frac{gC_{tW}}{4\pi^2}\frac{\sqrt{2}vm_t}{4\Lambda^2}
,\label{eq:U1}\\
V=&
-N_c\frac{gC_{tW}}{4\pi^2}\frac{\sqrt{2}vm_t}{\Lambda^2}\frac{m_W^2}{12m_t^2}
,\label{eq:V}\\
X=&
N_c\frac{gC_{tW}}{4\pi^2}\frac{\sqrt{2}vm_t}{\Lambda^2}\frac{5m_Z^2}{72m_t^2}s_Wc_W
.
\end{align}
where the result for $\hat U$ was already given in Eq.~(\ref{eq:Uloop}).  The one-loop contribution to the parameter $Y$ vanishes, and the one-loop contribution to the $W$ parameter is $-V$ [Eq.~(\ref{eq:V})].

In order to obtain constraints on $\hat{U}$, $V$ and $X$, we did a global fit using most major precision electroweak  measurements. These include the $Z$-pole data, the $W$-boson mass and width, DIS and atomic parity violation, and fermion pair production at LEP 2. The data and corresponding SM predictions can be found in
\cite{Nakamura:2010zzi,:2005ema,Schael:2006wu}.
The corrections to these observables from the seven oblique parameters
can be derived from the ``star'' formalism described in Ref.~\cite{Peskin:1991sw}.
We calculated the total $\chi^2$ as a function of the oblique parameters.
The central value for the fit is given by minimizing $\chi^2$, and the one-sigma bound is given by
$\chi^2-\chi^2_{\rm min}=1$. We let $\hat{S}$, $\hat{T}$, $W$ and $Y$ freely float and put constraints on the $\hat{U}$, $V$ and $X$ parameters.
We find three statistically independent combinations:
\begin{eqnarray}
0.46\hat{U}-0.46V+0.76X&=&-0.0013\pm0.0007,\label{eq:constrain1}\\
0.54\hat{U}-0.54V-0.65X&=&0.0000\pm0.0017,\label{eq:constrain2}\\
0.71\hat{U}+0.71V&=&-0.009\label{eq:constrain3}\pm0.030.
\end{eqnarray}
The most stringent constrain, Eq.~(\ref{eq:constrain1}), corresponds to
$\hat{U}-V+\frac{2s_Wc_W}{c_W^2-s_W^2}X$, which appears in the theoretical value of the $W$-boson mass:
\begin{align}
m_W^2=&m_W^2({\rm SM})
\left[
1-\frac{1}{c_W^2-s_W^2}\left(
2s_W^2\hat{S}-c_W^2\hat{T}-s_W^2W-s_W^2Y\right)
-\left(\hat{U}-V+\frac{2s_Wc_W}{c_W^2-s_W^2}X\right)
\right].
\end{align}
Combining Eqs.~(\ref{eq:U1}-\ref{eq:constrain1}) yields the constraint
\begin{equation}
\frac{C_{tW}}{\Lambda^2}=-3.4\pm2.0\ {\rm TeV}^{-2}.
\end{equation}
Including Eqs.~(\ref{eq:constrain2}) and (\ref{eq:constrain3})
gives a slightly better constraint,
\begin{equation}
\frac{C_{tW}}{\Lambda^2}=-2.8\pm1.8\ {\rm TeV}^{-2}.
\end{equation}
This constraint is weaker than the one given in
Eq.~(\ref{eq:CtW}),
but it is still comparable in precision to the constraints from direct measurements,
Eqs.~(\ref{eq:CDF}) and (\ref{eq:D0}).
It applies in more general situations than Eq.~(\ref{eq:CtW}),
as we only assume that the new physics is oblique (aside from $O_{tW}$).  The central value of $C_{tW}$ is nonzero at $1.5\sigma$, which indicates that the precision electroweak data have a slight preference for the presence of physics beyond the standard model.  

Constraints on the operator $O_{tW}$ may also be gleaned from $B$ physics.
This operator affects the branching ratio for $\bar B\to X_s\gamma$,
which is a loop-induced process.  It was found in Ref.~\cite{Grzadkowski:2008mf} that the contribution
from $O_{tW}$ is ultraviolet divergent.  Thus there must be a tree-level contribution from another
dimension-six operator, which masks the contribution from $O_{tW}$.  The operator $O_{tW}$ also affects $B-\bar B$ mixing, and it was found in Ref.~\cite{Drobnak:2011wj} that the contribution is ultraviolet finite, despite the fact that there are other dimension-six operators that contribute to this process at tree level.  Focusing only on $O_{tW}$, the constraint
\begin{equation}
\frac{C_{tW}}{\Lambda^2}=-0.06\pm 1.57 \ {\rm TeV}^{-2}.
\end{equation}
was obtained, which is comparable with the bounds from precision electroweak data [Eq.~(\ref{eq:CtW})] and top-quark decay [Eqs.~(\ref{eq:CDF}) and (\ref{eq:D0})].

We found that the indirect measurement of the coefficient of the operator
$O_{tW}$ from precision electroweak data is comparable in precision to
the direct measurement from top-quark decay.  The indirect measurement
will become more accurate with more precise electroweak measurements, in
particular of the $W$-boson mass.  The direct measurement will become
more accurate with more data from the Tevatron and the Large Hadron
Collider.  The direct measurement has the advantage that is affected, at
order $1/\Lambda^2$, only by the operator $O_{tW}$.  In contrast, there
are 4 operators (in the limit of $m_b\to 0$) that contribute to the $\hat U$ parameter at order
$1/\Lambda^2$, of which $O_{tW}$ is just one.  We will discuss this in a
companion paper on a global analysis of constraints on dimension-six
operators involving the top quark from precision electroweak data.

\section*{Acknowledgements}

We are grateful for correspondence with J.~Drobnak and J.~Serra. This material is based upon work
supported in part by the U.~S.~Department of Energy under contract
No.~DE-FG02-91ER40677 and the
National Science Foundation under Grant No.~0757889.

%

\end{document}